\documentclass[aps,prb,twocolumn]{revtex4}

\usepackage{epsfig}
\usepackage{graphicx}
\usepackage{amssymb}
\usepackage{mathrsfs}
\usepackage{dcolumn}
\usepackage{epstopdf}
\usepackage{amssymb}
\usepackage{amsfonts}

\newcommand{\om}{\omega}
\newcommand{\ra}{\rangle}

\newcommand{\be}{\begin{equation}} 
\newcommand{\ee}{\end{equation}} 
\newcommand{\bea}{\begin{eqnarray}} 
\newcommand{\eea}{\end{eqnarray}} 
\newcommand{\bqa}{\begin{eqnarray}}
\newcommand{\eqa}{\end{eqnarray}}
\newcommand{\bwt}{\begin{widetext}}
\newcommand{\ewt}{\end{widetext}}

\newcommand{\nn}{\nonumber \\}

\begin{document}

\title{A Novel Effect of Electron Spin Resonance on Electrical Resistivity}
\author{Navinder Singh}\email{navinder@prl.res.in}
\author{Luxmi Rani}\email{luxmiphyiitr@gmail.com}
\affiliation{Theoretical Physics Division, Physical Research Laboratory, Ahmedabad-380009, India.}

\begin{abstract}
We extend the well known phenomenon of magnetoresistance (extra resistivity of materials in transverse magnetic field) to a new and unexplored regime where in addition to a transverse magnetic field, a transverse AC field of resonant frequency is also applied. In a magnetic field, electron spin levels are Zeeman split. In a resonant AC field, we uncover a new channel of momentum relaxation in which electrons in upper Zeeman level can deexcite to lower Zeeman level by generating spin fluctuation excitation in the lattice (similar to what happens in Electron Spin Resonance (ESR) spectroscopy). An additional resistivity due to this novel mechanism is predicted in which momentum randomization of Zeeman split electrons happen via bosonic excitations (spin fluctuations). An order of magnitude of this additional resistivity is calculated. The whole work is based upon an extension of Einstein's derivation of equilibrium Planckian  formula to near equilibrium systems.
\end{abstract}

\pacs{.......}

\maketitle

\begin{figure}[htbp!]
       \centering
        \includegraphics[angle=0,width=5.5cm]{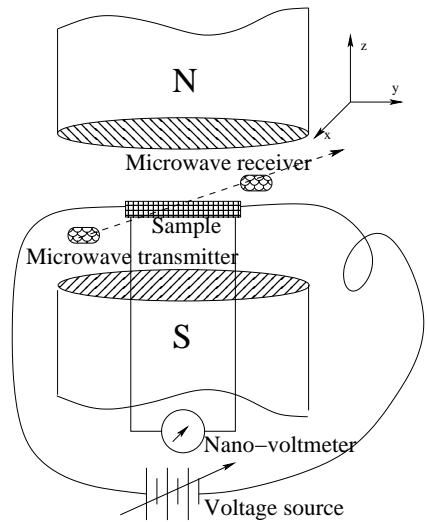}
         \caption{Schematic diagram of the experimental set-up.}
          \label{fig:figa}
 \end{figure}

{\it Introduction}: The phenomenon of magnetoresistance is well studied\cite{magneto}. Magnetoresistance is an extra resistance of materials in transverse magnetic field. We uncover a hitherto neglected mechanism of resistivity and investigate, for the first time, another ``extra'' resistance in some specific materials that will appear when alongwith the transverse magnetic field, an AC field of resonant frequency is applied (Fig.\ref{fig:figa}). In the figure a sample is placed in a uniform magnetic field which is directed in Z-direction (say). A current is impressed through the sample in the Y-direction. This is the standard set up for magnetoresistance measurements. We introduce another dimension to this experimental set-up (which seems to bring far reaching results). We introduce a resonant AC field along the X-direction (Fig.\ref{fig:figa}). Thus, a current is passed through the the sample which is placed in a criss-crossed static magnetic field and an AC resonant field of appropriate frequency. The frequency $\omega$ of the AC field is chosen such that $\hbar\omega =\mu_B H_{eff}$ where $H_{eff}$ is the effective field ``seen" by the conduction electrons in the sample. It is equal to the sum of the external field (H) and any induced internal magnetic field ($H_{int}$) in the sample. The condition $\hbar\omega =\mu_B H_{eff}$ says that the energy of the photon $\hbar\omega$ is resonant with Zeeman splitting of conduction electron energy. This condition is the same as applied in ESR spectroscopy.
 
 In the following development we will show that this kind of experimental arrangement leads to an ``extra'' resistance of the sample, and it is related to the microscopic parameters of the material in a fundamental way. Measurement of it can lead to the determination of those parameters. The mechanism involved is novel one and to the best of our knowledge, it has never been discussed before. We also determine the order of magnitude of the proposed effect and find that it is well within the current measurements capabilities. We dub this ``extra" resistance as Magneto-electro-resistance (generalizing  the concept of magnetoresistance). It is to be noted that our proposed effect is fundamentally different from Electrically Detected Magnetic Resonance (EDMR) phenomenon. In EDMR, spin states of electrons in semiconductors on donor impurities are flipped by on external microwave source, and thereby enabling them to recombine with conduction holes. This recombination remove charge carriers from the conduction band hence increasing its resistivity\cite{Ghosh,Vlasenko,Graeff,Fedorych,Stegner,Boehme,DRM, Lo,CB, Zhu,Lee,SYL, Hoehne,Franke,Dreher,Keevers,Behrends}.
 The field which we wish to open is different one and a hybrid of magnetoresistance phenomenon and NMR spectroscopy, and can lead to novel electronic transport measurements.

We now proceed to the formulation of the effect and explanation of the mechanism. We have the following physical picture:

  \begin{itemize}
  \item The sample is placed in a magnetic field of strength H which is directed along Z-direction.
  \item In transverse direction, an AC field of frequency $``\omega"$ is applied, such that there is an energy density $u(\omega)$ of the field in the sample. 
  \item Along X-direction, current I is made to flow by an external field $E$. 
  \end{itemize}

 The phenomenological formulation which we put forward goes like this. Under the action of magnetic field conduction electrons turns into ``Itinerant two-level systems" due to Zeeman splitting. And these Itinerant Two-Level Systems (ITLS) drift when external electric field is applied. If a given ITLS is in an upper excited state, it can get rid off an energy $\delta E = E_2 - E_1$ by three process: (1) by spontaneously giving off a radiation quantum, or (2) by an excitation of a spin fluctuation in the lattice, and (3) by stimulated emission. Similarly, excitation from lower Zeeman level to upper Zeeman level can happen as discussed below. We wish to sketch a calculation along the lines of Einstein's derivation of Planck's formula. 
 
 \begin{figure}[htbp!]
       \centering
        \includegraphics[angle=0,width=7.5cm]{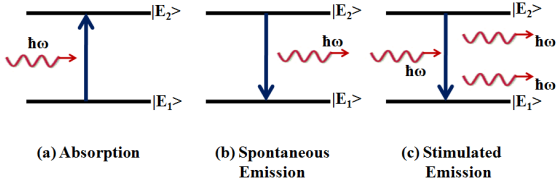}
         \caption{Schematic diagram of the interaction of an itinerant two level system with field and lattice.}
          \label{fig:figEB}
 \end{figure}

  \textbf{Case-I}: Absorption ($|E_1\rangle-|E_2\rangle$). The rate at which electrons excite from lower Zeeman level $|E_1\rangle$ to upper Zeeman level $|E_2\rangle$ is given by:
 \bea
 \frac{dn_{12}}{dt} = B_{12} u(\omega) n_1 + \alpha \frac{1}{\tau_{spin}}n_1.
 \label{eq:ab}
  \eea
 Here, $n_1$ is the number of Zeeman split electrons per unit volume which are in $|E_1\rangle$ or ground state (Fig.\ref{fig:figEB}). The first term on the R.H.S. of the above equation is the standard Einstein term which says that excitation rate is proportional both to the strength of radiation density $u(\om)$ and to the number of electrons in the lower level. The proportionality constant is $B_{12}$ (Einstein's $B$ coefficient). In the second term (which we phenomenologically introduce), $1/\tau_{spin}$ represents the collision rate with spin fluctuation in the material\footnote{In a transport setting electron-phonon interaction is important in simple metals; in magnetic metals electron-spin scattering becomes an important interaction like in $MnSi, ZrZn_2$. }. $\alpha$ is the fraction of collisions that leads to $|E_1\ra \rightarrow |E_2\rangle$ while 1-$\alpha$ does not lead to $|E_1\rangle\rightarrow |E_2\rangle$, in other words it measures the success rate of upward transition.

 \textbf{Case-II}: An electron in the upper Zeeman level can de-excite by the following process: (1) spontaneous emission:
  \bea
 \frac{dn_{21}}{dt} = A_{21} n_2.
  \eea
  Here, $n_2$ is the number of Zeeman split electrons per unit volume in upper $|E_2\rangle$ state ($n_1 +n_2 =1$).  $A_{21}$ is the Einstein coefficient of spontaneous emission. (2) Stimulated emission and ``spin fluctuation'' generation
  \bea
 \frac{dn_{21}}{dt} = B_{21} u(\omega) n_2 + \gamma \frac{1}{\tau_{spin}}n_2. 
 \label{eq:st}
  \eea
Where, $0< \gamma <1$ measures the success rate of the downward transition due to spin fluctuation generation (a spin fluctuation is a bosonic excitation with spin quantum number one, like photon. An electron de-exciting from upper level to lower level suffer a change of spin quantum number by one unit, and the generated bosonic excitation also has spin quantum number one. Thus the total spin is conserved in the process, as it should). $B_{21}$ is the Einstein coefficient of induced (or stimulated) emission. In equations (\ref{eq:ab}) and (\ref{eq:st}), we introduce additional new terms $\alpha \frac{1}{\tau_{spin}}n_1$ and $\gamma \frac{1}{\tau_{spin}}n_2$ which takes care of transitions induced by spin fluctuation absorption and generation, respectively. Rest of the terms are standard. 

In a steady state situation: Absorption $\equiv$ Emission:  
  \bea
  B_{12} u(\omega)n_1 + \frac{\alpha }{\tau_{spin}}n_1
  &=& A_{21}n_2 + B_{21} u(\omega) n_2 + \frac{\gamma }{\tau_{spin}}n_2,\nn
  \eea
  or,
  \bea
   \frac{B_{12} u(\omega) + \frac{\alpha }{\tau_{spin}}}{A_{21} + B_{21}u(\omega)+  \frac{\gamma}{\tau_{spin}}}
   &=& \frac{n_2}{n_1} = \frac{e^{-\beta E_2}}{e^{-\beta E_1}}\simeq e^{-\beta \omega \hbar},
\eea
where $\hbar \omega = E_2 -E_1$ and $\beta = \frac{1}{k_B T}$. $k_B$ is the Boltzmann constant. We assume that the system remains in a near equilibrium state (a quantitative analysis of this is presented below). On simplifying the above equation we have
\bea
 \frac{1}{\tau_{spin}}(\alpha e^{\beta \hbar\omega} -\gamma) = A_{21}+ B_{21}u(\omega)- B_{12}u(\omega)e^{\beta \hbar \omega}.\nn 
 \label{eq:5}
\eea
At equilibrium, one must have $\alpha e^{\beta \hbar\omega} = \gamma $. Thus, R.H.S of above equation is zero. Equation (\ref{eq:5}) leads to 
\bea 
  u(\omega)= \frac{A_{21}}{B_{12}e^{\beta \omega \hbar}-B_{21}}.
   \label{eq:u(w)}
  \eea
  On applying the standard boundary conditions into  equation(\ref{eq:u(w)}); where energy density $u(\omega)$ must tends to infinity ($u(\omega)\rightarrow \infty$), when $T \rightarrow \infty$; leads to  $ B_{12} =B_{21} =B$. Thus
 \bea 
  u(\omega)= \frac{A}{B(e^{\beta \omega \hbar}-1)},
 \eea
  which is the Planck's law and the ratio $A/B$  is computed\cite{ele} by the Raleigh-Zean limit $\omega \rightarrow 0$. 
 
 The situation in which we are interested is a near equilibrium situation. 
 Near equilibrium: $\alpha e^{\beta \hbar\omega} \lessapprox \gamma $ (this condition is discussed in the last section), then from equation (\ref{eq:5}) we have
\be
 \frac{1}{\tau_{spin}} = Bu(\omega) \bigg(\frac{e^{\beta \hbar \omega}-1}{ \gamma -\alpha e^{\beta \hbar \omega}}\bigg)-A\bigg(\frac{1}{\gamma -\alpha e^{\beta \hbar \omega}}\bigg).    
 \label{eq:imp}
\ee
This is an important result in which spin fluctuation scattering rate is expressed in terms of phenomenon parameters $\alpha$, $\gamma$, $u(\omega)$, $A$, and $B$.

{\it Electro-magneto Resistance using the Drude theory}: To calculate the extra resistance due to spin fluctuation generation and absorption we proceed with the simple Drude theory of momentum relaxation. This is similar to the solution of the Boltzmann equation under relaxation time approximation\cite{nav}. The Drude theory is characterized by a constant momentum relaxation rate $1/\tau$. If $\overrightarrow{\rm P}$ is an average momentum of an electron, then the equation of motion under various relaxation processes is given by: 

 \bea
 \frac{d\overrightarrow{\rm P}}{dt}= -e\overrightarrow{\rm E} -\frac{1}{\tau_{others}}\overrightarrow{\rm P}-\frac{\alpha}{\tau_{spin}}n_{1}\overrightarrow{\rm P} -\frac{\gamma}{\tau_{spin}}n_2 \overrightarrow{\rm P}.\nn
 \label{eq:eom}
  \eea
Here $\frac{1}{\tau_{others}}$ is the standard Drude relaxation rate due to impurity scattering. We introduce two more relaxation terms: $-\frac{\alpha}{\tau_{spin}}n_{1}\overrightarrow{\rm P}$ and $ -\frac{\gamma}{\tau_{spin}}n_2 \overrightarrow{\rm P}$. The first term denotes momentum randomization of an electron when it absorbs a spin fluctuation quantum (as explained below equation (\ref{eq:ab})). The second term (latter term) denotes the momentum randomization of a drifting electron when it generates a spin fluctuation quantum and thereby de-excites from the higher Zeeman level to the lower Zeeman level. The momentum randomization happens as the direction of these excitations is random ({\it much like what happens in the case of spontaneous emission, in which an atom suffer random kicks due to spontaneously emitted photons}). As we are considering a near equilibrium situation we have $\frac{n_2}{n_1}= e^{-\beta \hbar \omega}$, where $n_1=\frac{e^{\beta \hbar \omega}}{e^{\beta \hbar \omega}+1}$, $n_2=\frac{1}{e^{\beta \hbar \omega}+1},~n_1+n_2 =1$. The equation of motion (equation (\ref{eq:eom})) takes the form:
\bea
 \frac{d\overrightarrow{\rm P}}{dt}= -e\overrightarrow{\rm E} -\overrightarrow{\rm P}\bigg(\frac{1}{\tau_{others}}+\frac{1}{\tau_{spin}}\bigg(\frac{\alpha e^{\beta \hbar \omega}+ \gamma}{e^{\beta \hbar \omega} + 1}\bigg)\bigg)
\eea
In the Ohmic regime $\overrightarrow{\rm J} =  \sigma \overrightarrow{\rm E} = -e n\overrightarrow{\rm V}$,  and  $\overrightarrow{\rm P} =  m\overrightarrow{\rm V}$. Thus, the resistivity ($\rho = 1/\sigma$) is 
\bea
\rho = \frac{m}{ne^2}\bigg(\frac{1}{\tau_{other}}+ \frac{1}{\tau_{spin}}\bigg(\frac{\alpha e^{\beta \hbar \omega}+ \gamma}{e^{\beta \hbar \omega} + 1}\bigg)\bigg)
\eea
Here, the second term (the novel term) is additional resistivity with
\bea
\frac{1}{\tau_{additional}}&=&\bigg(\frac{\gamma +\alpha e^{\beta \hbar \omega}}{\gamma-\alpha e^{\beta \hbar \omega}}\bigg)\times \nn
&&\bigg(Bu(\omega)\tanh\bigg(\frac{1}{2}\beta \hbar\omega \bigg) - \frac{A}{e^{\beta \hbar \omega}+1}\bigg)\nn\,
\label{eq:adscat}
\eea
where we used equation (\ref{eq:imp}). Thus the additional resistivity is given by

\bea
\rho_{additional} &=&  \frac{m}{n e^2}\frac{1}{\tau_{additional}}= \frac{m}{n e^2}\bigg(\frac{\gamma +\alpha e^{\beta \hbar \omega}}{\gamma-\alpha e^{\beta \hbar \omega}}\bigg) \nn
&&\times\bigg(Bu(\omega)\tanh\bigg(\frac{1}{2}\beta \hbar\omega \bigg) - \frac{A}{e^{\beta \hbar \omega}+1}\bigg)\nn.
\eea
This is the central result of the present investigation. Thus there must be an ``extra'' resistivity solely coming from the excitation and absorption of spin fluctuations in the lattice whose quantum is equal to the separation between the Zeeman levels of the drifting electron. {\it The importance of this new effect is that it can be externally tuned by controlling the Zeeman splitting through external magnetic field.} Spontaneous rate in condensed matter systems is generally weak. So the result simplifies to:
\bea
\rho_{additional}&=& \frac{m}{n e^2} Bu(\omega)\bigg(\frac{\gamma +\alpha e^{\beta \hbar \omega}}{\gamma-\alpha e^{\beta \hbar \omega}}\bigg)
tanh\bigg(\frac{1}{2}\beta \hbar\omega \bigg).\nn\
\eea

{\it Typical values and the order of magnitude of the effect}: To roughly estimate of the order of magnitude of the effect we take values for a typical metallic sample which is placed in a magnetic field of $1$ Tesla (for example) and at temperature of $100~K$ $(\mu_B H = \hbar\om)$. Let us take $B u(\om) \simeq 10^{10}~Hz$ (refer to footnote\footnote{From Fermi golden rule, it can be roughly approximated as: $\frac{1}{\tau}\simeq\frac{2\pi}{\hbar}|V|^2 \rho(E)$, where $\rho(E)\sim a^{-3}$, $a$ is the lattice constant ($a \sim 1 \AA$) and V is in sub-$\mu$eV range. Thus $\frac{1}{\tau}\sim 10^{10}$ in per sec.}). The spontaneous transition rate of a Zeeman spit electron from upper level to lower level in  metals is of the order of $MHz$\footnote{This can be inferred from the ESR linewidths which are in MHz range. The linewidths have contributions from interactions in the material along with spontaneous rate.}. Thus, we can safely neglect its contribution to additional scattering rate. We assume that $\frac{\gamma}{\alpha}\simeq 1.01$. The regime corresponding to this ratio is investigated below. With this ratio one is not far from equilibrium as shown below. Having this input of the physical parameters, the additional scattering rate (from equation (\ref{eq:adscat})) turns out to be: $\frac{1}{\tau_{additional}} \simeq 10^{10}(Hz)$, where $\tanh(x)\sim x$ for small $x$. The extra resistivity due to this is given by $\rho_{extra} = \frac{m}{n e^2}(1/\tau_{add}) \simeq 10^{-9} \Omega.~meters$ (we take $n= 10^{26}$ per meter cube (typical electron number density in metals)). 

If we have a sample of a size $4~mm \times 4~mm\times 4~mm$, then the extra resistance of it will be $\sim 1 micro \Omega$. This should not be difficult to measure. If a current of $100~mA$ is made to flow in the sample, then change in the voltage drop (due to change in the resistance $\Delta V = I \Delta R$) will be $100 nano-volts$, which  again can be detected with a sensitive nano-voltmeter.

{\it The operational regime}: We now investigate what is meant by $\gamma \gtrapprox \alpha e^{\beta\hbar\om}$ (the near equilibrium condition)\cite{nmr}. We neglect the weaker spontaneous emission term. We consider the two cases separately:

Case (1) External fields only: 
 \bea
 \frac{dn_1}{dt}= n_2 B_{21} u(\omega) -n_1 B_{12}u(\omega).
  \eea
Here, let us set $n=n_1 -n_2$  and $N=n_1 + n_2$. Then, we have
\bea
\frac{1}{2}\frac{dn}{dt}=-n B u(\om).
\eea
After simplification, we get the relation;
\bea
n(t) =n(0)e^{-2 B u(\om) t}.
\eea
Power absorbed by the sample from the field:
\bea
P_{absorbed}= n_1 B u(\om) \hbar \omega -n_2  B u(\om) \hbar \omega
&=& n B u(\om)\hbar\omega.\nn\
\eea
Case (2): Spin-fluctuation scattering only:
\bea
\frac{dn_1}{dt}= n_2 \left(\frac{\gamma}{\tau_s}\right) -n_1 \left(\frac{\alpha}{\tau_s}\right)
\label{eq:A1}
\eea
In the equilibrium state: $\frac{dn_1}{dt}=0$, thus
\bea
\frac{n_2}{n_1}=\frac{\alpha}{\gamma}=e^{-\beta \hbar\omega},
\eea
or, $\alpha e^{\beta \hbar\omega} =\gamma$ is the equilibrium condition.

From equation (\ref{eq:A1}), we have
\bea
T_1 \frac{dn}{dt} =n_0 -n,
\eea
where, $T_1= \frac{\tau_s}{\alpha + \gamma}$ (as usually defined in the NMR spectroscopy) and $n_0 = N \frac{\gamma - \alpha}{\gamma + \alpha}$.
\bea
 n(t)= n_0 (1- e^{-t/T1}).
\eea
When both are acting together in a steady state situation
\be
\bigg(\frac{dn}{dt}\bigg)_{spin-fluc}  + \bigg(\frac{dn}{dt}\bigg)_{field} =0,
\ee
thus
\be
n(t) = \frac{n_0}{1+ 2 B u(\om)T_1}.
\ee

There are two possible cases:

Case A: If $2 B u(\om) T_1 \ll 1$ or $T_1 \ll \frac{1}{2 B u(\om)}$ $(\frac{\alpha + \gamma}{\tau_s}\gg 2Bu(\omega))$. $n(t) \simeq n_0$. This means Spin scattering is fast, system stays in equilibrium.  

Case B: If $2 B u(\om) T_1 \gg 1$ or $(\frac{\alpha + \gamma}{\tau_s}\ll 2Bu(\omega))$. This means strong field and system stays away from equilibrium.

{\it Power Absorbed}:
\bea
P_{ab} = n B u(\om) \hbar \omega = \frac{n_0 \hbar \omega B u(\om)}{1 + 2 B u(\om)T_1}.
\eea

\begin{figure}[htbp!]
        \centering
         \includegraphics[angle=0,width=6cm]{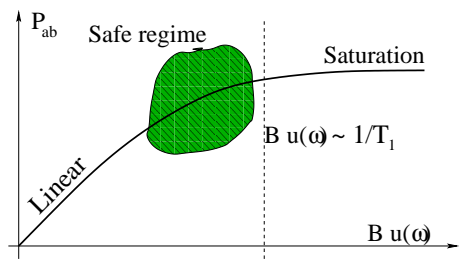}
          \caption{Schematic depiction of the intermediate regime (green shaded area) in which the system is to be forced by tunning $u(\om)$.}
           \label{regime}
\end{figure}

The regime which we consider is the intermediate one (figure (\ref{regime})). In the example numerical case mentioned above, the power absorbed is in micro-watts.

{\it Conclusion} The mechanism of spin wave generation when a Zeeman split electron makes a downward transition from upper Zeeman level to lower Zeeman level leads to random kicking of the conduction electron. This random kicking is the source of momentum randomization of the electron and leads to extra resistivity which we uncover for the first time. Similarly, the absorption of spin waves leads to random kicking and extra resistance. We calculate an order of magnitude of this effect and we find that it is well within the detectable limits.

{\it Acknowledgement}: We thank to Rajesh Kumar Kushawaha, Prashant Kumar, K. P. Subramanian and Gopalakrishnan Balasubramanian for useful suggestions and discussions.



\begin{thebibliography}{201}

\bibitem{magneto} A. B. Pippard, {\it Magnetoresistance in metals}, CUP, Cambridge, UK (1989); R. G. Chambers, {\it Electrons in Metals and Semiconductors}, Springer, Netherlands (1990).
\bibitem{Ghosh} R. N. Ghosh and R. H. Silsbee, {\it Spin-spin scattering in a silicon two- dimensional electron gas}, Phys. Rev. B \textbf{46}, 12508 (1992).

\bibitem{Vlasenko} L.S. Vlasenko, Yu. V. Martynov,T. Gregorkiewicz and C.A. J. Ammerlaan, {\it Electron paramagnetic resonance versus spin-dependent recombination: Excited triplet states of structural defects in irradiated silicon}, Phys. Rev. B \textbf{52}, 1149 (1995).

\bibitem{Graeff} C. F. O. Graeff, M. S. Brandt, M. Stutzmann, M. Holzmann, and G. Abstreiter, F. Sch\"{a}ffler, {\it Electrically detected magnetic resonance of two-dimensional electron gases in Si/SiGe heterostructures}, Phys. Rev. B \textbf{59}, 242 (1999).

 \bibitem{Fedorych} O. M. Fedorych, Z. Wilamowski, W. Jantsch and J. Sadowski, {\it Electrically detected magnetic resonance}, Acta Physica Polonica A \textbf{105}, 591 (2004).
 \bibitem{Boehme} C. Boehme and H. Malissa, {\it Electrically detected magnetic resonance spectroscopy}, eMagRes \textbf{6}, 83 (2017).
 \bibitem{Stegner} A. R. Stegner, C. Boehme, H. Huebl, M. Stutzmann, K. Lips and M. S. Brandt, {\it Electrical detection of coherent $^{31}$P spin quantum states}, Nature Phys. \textbf{2}, 835 (2006).
\bibitem{DRM} D. R. McCamey,  H. Huebl, M. S. Brandt, W. D. Hutchison, J. C. McCallum, R. G. Clark, and A. R. Hamilton, {\it Electrically detected magnetic resonance in ion-implanted Si:P nanostructures}, Appl. Phys. Lett. \textbf{89}, 182115 (2006).
\bibitem{Lo} C. C. Lo, V. Lang, R. E. George, J. J. L. Morton, A. M. Tyryshkin, S. A. Lyon, J. Bokor, and T. Schenkel, {\it Electrically detected magnetic resonance of neutral donors interacting with a two-dimensional electron gas}, Phys. Rev. Lett. \textbf{106},
207601 (2011).
\bibitem{CB} C. Boehme and K. Lips, {\it Theory of time-domain measurement of spin-dependent recombination with pulsed electrically detected magnetic resonance}, Phys. Rev.B
68, 245105 (2003).
\bibitem{Zhu} L. Zhu, K. J. van Schooten, M. L. Guy and C. Ramanathan, {\it Optical dependence of electrically-detected magnetic resonance in lightly-doped Si:P devices}, Phys. Rev. Applied \textbf{7}, 064028 (2017).
\bibitem{Lee} S.-Y. Lee, S. Paik, D.R. McCamey and C. Boehme, {\it Modulation frequency dependence of continuous-wave optically/electrically detected magnetic resonance}, Phys. Rev. B \textbf{86}, 115204 (2012).
\bibitem{SYL} S.-Y. Lee, S. Paik, D.R. McCamey and C. Boehme, {\it Modulation frequency dependence of continuous-wave optically/electrically detected magnetic resonance}, Phys. Rev. B \textbf{86}, 115204 (2012).
 \bibitem{Hoehne} F. Hoehne, L. Dreher, J. Behrends, M. Fehr, H. Huebl, K. Lips, A. Schnegg, M. Suckert, M. Stutzmann and M. S. Brandt, {\it Lock-in detection for pulsed electrically detected magnetic resonance}, Review of Scientific Instruments \textbf{83}, 043907 (2012).
  \bibitem{Franke} D. P. Franke, F. Hoehne, L. S. Vlasenko, K. M. Itoh and M. S. Brandt, {\it Spin-dependent recombination involving oxygen-vacancy complexes in silicon}, Phys. Rev. B \textbf{89}, 195207 (2014).
   \bibitem{Dreher}L. Dreher, F. Hoehne, H. Morishita, H. Huebl, M. Stutzmann, K. M. Itoh and M. S. Brandt, {\it Pulsed low-field electrically detected magnetic resonance}, Phys. Rev. B \textbf{91}, 075314 (2015).
  \bibitem{Keevers}T. L. Keevers, W. J. Baker and D. R. McCamey, {\it Theory of exciton-polaron complexes in pulsed electrically detected magnetic resonance}, Phys. Rev. B \textbf{91}, 205206 (2015).
  \bibitem{Behrends} J. Behrends, A. Schnegg, K. Lips, E. A. Thomsen, A. K. Pandey, I. D. W. Samuel  and D. J. Keeble, {\it Bipolaron formation in organic solar cells observed by pulsed electrically detected magnetic resonance}, Phys. Rev. Lett. \textbf{105}, 176601 (2010)

\bibitem{ele} H. Haken and H. C. Wolf, {\it The Physics of Atoms and Quanta}, Springer-Verlag, Berlin, Heidelberg (2004); R. Eisberg and R. Resnick, {\it Quantum Physics}, John Wiley and Sons,  New York (1985).

\bibitem{nav} N. Singh, {\it Electronic Transport Theories: From Weakly to Strongly Correlated Materials}, CRC Press, Taylor and Francis group, LLC, UK (2017).

\bibitem{nmr} C. P. Slichter, {\it Principles of Magnetic Resonance}, Springer-Verlag, Berlin, Heidelberg (1990).





 

 



 








\end{thebibliography}
\end{document}